\documentclass[prb,amsmath,amssymb,
reprint,%
]{revtex4-2}

\usepackage{graphicx}
\usepackage{dcolumn}
\usepackage{bm}

\usepackage[utf8]{inputenc}
\usepackage[T1]{fontenc}
\usepackage{mathptmx}
\usepackage{xcolor}

\usepackage[colorlinks=true,linkcolor=blue,citecolor=blue,urlcolor=blue,allbordercolors=white]{hyperref} 

\begin{document}

\title{Signaling oscillations: molecular mechanisms and functional roles}

\author{Pablo Casani-Galdon}
\email{pablo.casani@upf.edu}
\affiliation{Department of Medicine and Life Sciences, Universitat Pompeu Fabra,
Barcelona Biomedical Research Park, Dr. Aiguader 88, 08003 Barcelona, Spain}

\author{Jordi Garcia-Ojalvo}
\email{jordi.g.ojalvo@upf.edu}
\affiliation{Department of Medicine and Life Sciences, Universitat Pompeu Fabra,
Barcelona Biomedical Research Park, Dr. Aiguader 88, 08003 Barcelona, Spain}

\begin{abstract}
Mounting evidence shows that oscillatory activity is widespread in cell signaling.
Here we review some of this recent evidence, focusing on both the molecular mechanisms that potentially underlie such dynamical behavior, and the potential advantages that signaling oscillations might have in cell function.
The biological processes considered include tissue maintenance, embryonic development and wound healing.
With the aid of mathematical modeling, we show that a common principle, namely delayed negative feedback, underpins this wide variety of phenomena.
\end{abstract}

\maketitle

\section{Introduction}

Signaling allows cells to sense, integrate and respond to cues from their environment, and most importantly, to coordinate their behavior with neighboring cells, thereby enabling cellular populations to self-organize in space and time.
The textbook view of cellular signaling usually considers, either explicitly or implicitly, that the signals received by cells are constant in time.
However, the cellular environment is commonly dynamical \emph{in vivo}, and such time-dependent character is bound to have influenced the evolution of signaling circuits.

One way in which signaling circuits can cope with the dynamical, and usually poorly predictable, external signals that cells receive, is by being dynamical themselves \citep{henrique2019mechanisms,Gabalda-Sagarra:2018uv}.
In agreement with this expectation, examples of oscillatory and pulsatile signaling have begun to be uncovered in recent years \citep{cheong2010oscillatory,levine2013}.
Early instances of oscillatory signaling were reported in the last decades of the past century, in processes including glycolysis in muscle and yeast cells, and cAMP signaling in \textit{Dyctiostelium} \cite{sel1968self, pye1966sustained, goldbeter1997biochemical}.

Two questions arise in this context.
First, what are the molecular mechanisms underlying such self-sustained dynamical behavior?
Second, what are the roles of signaling oscillations in cells?
Regarding the latter question, the function of dynamical signaling is still unclear in many cases. 
In some situations, oscillations could arise as a byproduct of adaptation, with no relevant biological function \cite{cheong2010oscillatory}. 
In other cases, however, oscillations have been found to be crucial for the proper operation of cells, such as in the response of p53 to DNA damage \citep{Purvis:2013tf}, the maintenance of the neural progenitor state \citep{Imayoshi2013}, the segmentation of vertebrates \cite{oates2012patterning, sonnen2018modulation}, and the growth of bacterial biofilms \cite{Martinez-Corral:2019vg}. 

\begin{figure}[htbp]
    \centering
    \includegraphics[scale=0.6]{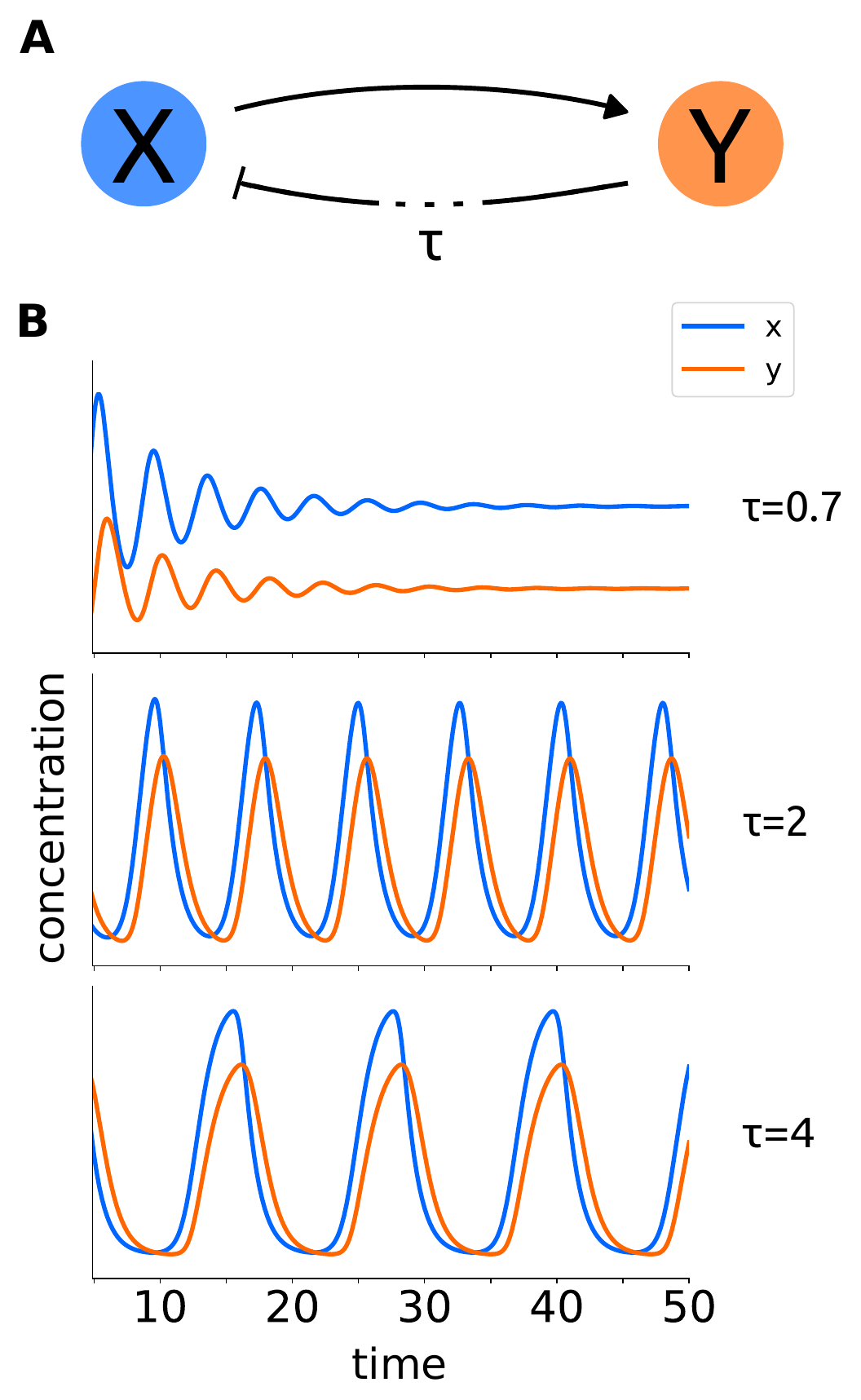}
    \caption{Oscillations can arise from delayed negative feedback.
    (\textbf{A}) Schematic representation of the circuit.
    The solid line represents a direct interaction and the dashed line represent a delayed interaction with time lag $\tau$.
    (\textbf{B}) Simulation of the system \ref{eq:sys1} for increasing values of the time lag $\tau$.
    The values of the rest of parameters, together with the simulation code in Julia, can be found in
    \href{https://github.com/dsb-lab/SignalingOscillators}{https://github.com/dsb-lab/SignalingOscillators}.}	
    \label{fig:dde}
\end{figure}

Beyond their functional role, the existence of oscillations in certain signaling systems may help us to identify and dissect the molecular circuits underlying these cellular processes: while a biochemical system can reach a steady state in many different ways, the number of potential circuits and parameter values that can produce oscillations with the right properties (period, amplitude, and response to perturbations, for instance) is much smaller \citep{Kirk:2008ws}.
Indeed, a relatively small number of principles underlying the emergence of self-sustained oscillations have been identified \citep{novak2008design}.
The most fundamental of these principles is delayed negative feedback \citep{Martinez-Corral:2018uc}.
Simple mathematical models show that when a signaling pathway inhibits itself with a time lag (Fig.~\ref{fig:dde}A), large enough delays trigger self-sustained oscillations, as identified in pioneering models of the segmentation clock and other recurrent pathways \citep{monk2003oscillatory,lewis2003autoinhibition}.
Mathematically, this can be represented for instance by the following system of delayed-differential equations:
\begin{subequations}\label{eq:sys1}
\begin{align}
	\frac{dX}{dt}&=\frac{\beta_x K_r^n }{K_r^n + Y^n(t - \tau)} - \delta_x X \label{eq:sys1a}\\
	\frac{dY}{dt}&=\frac{\beta_y X^m }{K_a^m + X^m} - \delta_y Y \label{eq:sys1b}
\end{align}
\end{subequations}
Here the species $X$ activates $Y$ in a sigmoidal manner, with threshold $K_a$, and $Y$ inhibits $X$ (also in a sigmoidal manner, with threshold $K_r$) after a time delay $\tau$, represented by the explicit dependence of $Y$ on $t-\tau$ in the first term on the right hand side of Eq.~(\ref{eq:sys1a}) (all other instances of $X$ and $Y$ in the equations above are computed at time $t$, reflecting instantaneous processes).
As shown in Fig. \ref{fig:dde}B, simulating the system of equations (\ref{eq:sys1}) leads to self-sustained oscillations for large enough $\tau$.

In model (\ref{eq:sys1}), $X$ could represent an mRNA species that is translated into its own protein $Y$, although other molecular interpretations (such as $X$ and $Y$ being two different proteins in a recurrent pathway) are possible.
The time lag $\tau$ can also affect the other interaction in the circuit (from $X$ to $Y$), or be distributed among the two \citep{matsuda2020species}.
The functional forms of the various terms in Eqs.~(\ref{eq:sys1}) can also be different, as long as the total nonlinearity in the feedback loop is large enough \citep{Murray:2002ur}.
There is thus substantial freedom in how the delayed negative feedback can be implemented, and indeed evolution has come across this design principle in many different ways.
In what follows we review three different instances of oscillatory signaling circuits that implement negative feedback in various forms, with the time delay arising in correspondingly distinct manners.
We discuss the molecular mechanisms underlying the different circuits, as well as the biological functions that oscillations have in each case.
All simulations shown in this review have been generated with custom-made Julia code that can be obtained from \href{https://github.com/dsb-lab/SignalingOscillators}{https://github.com/dsb-lab/SignalingOscillators}.

\section{Notch oscillations in tissue maintenance}

The Notch signaling system is a classical juxtacrine pathway involved in the formation and maintenance of most tissues, among many other cellular functions \cite{bray2016notch,Guisoni:2017wb}. 
Even though this pathway is involved in different lineage decisions, its core circuitry is highly conserved \cite{guruharsha2012notch}. 
The versatility of Notch signaling relies on its multiple levels of control \citep{nakayama2008fgf,bernard2010specificity,nandagopal2019cis}, its high sensitivity to perturbations \cite{henrique2019mechanisms}, and its ability to exhibit different responses in the presence of different ligands \citep{nandagopal2018dynamic}.
Some of these properties arise from, or are enabled by, the dynamical character of the pathway \citep{henrique2019mechanisms,nandagopal2018dynamic}.

\begin{figure*}[htb]
\centering
\includegraphics[scale=0.4]{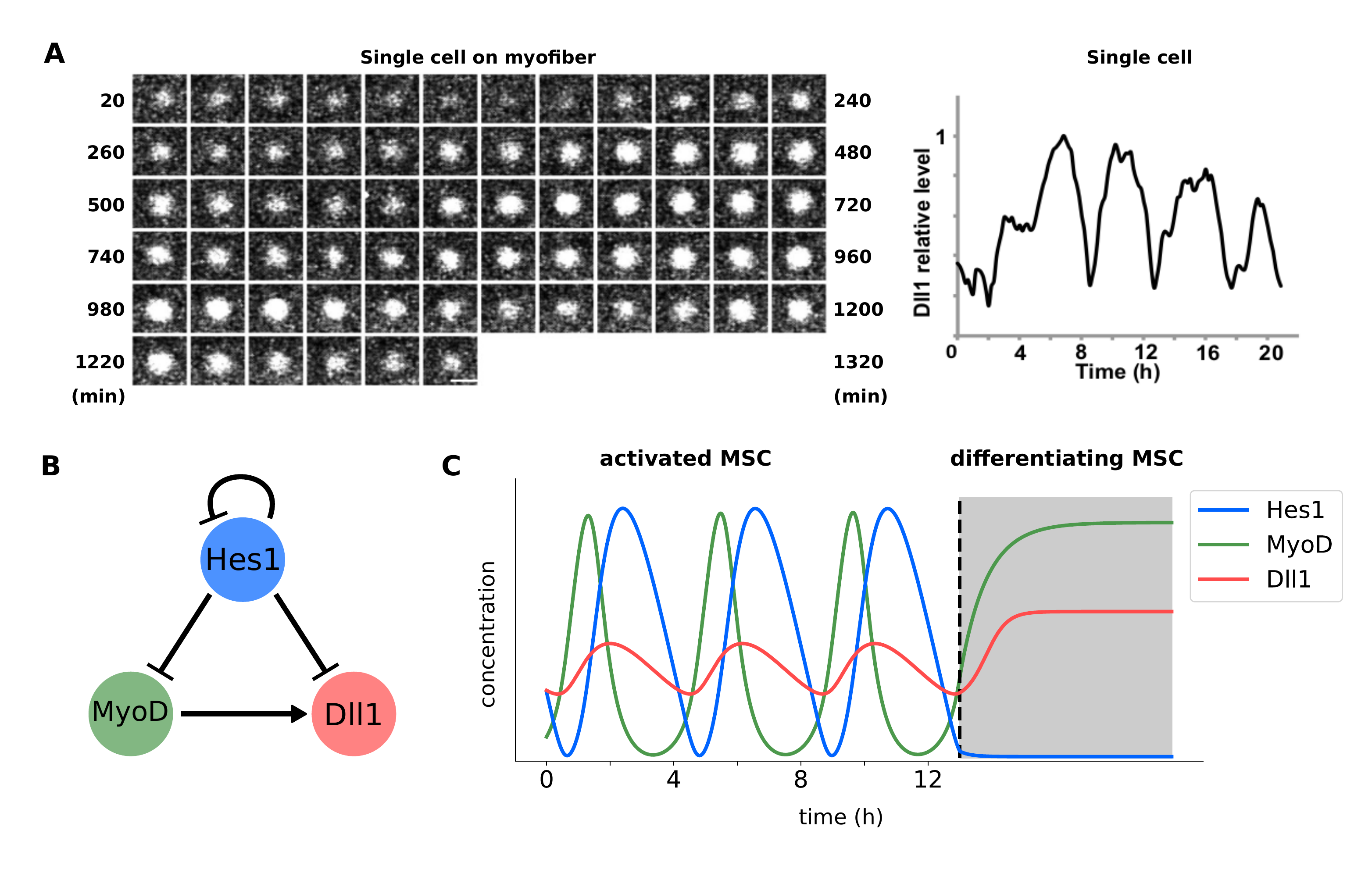}
\caption{Oscillations in Notch signaling. 
(\textbf{A}) Dll1 oscillates in activated muscular stem cells. (\textbf{B}) Schematic representation of the whole Hes, MyoD and Dll1 circuit. (\textbf{C}) Time traces of Hes1, MyoD, Dll1 simulated with Eqs.~(\ref{eq:sys2}) and its downstream elements as presented in \cite{zhang2021oscillations}. The vertical dashed line represents the moment of the commitment and the end of the oscillations.
Adapted from \cite{zhang2021oscillations}. The simulations in panel (\textbf{C}) were generated with the Julia code posted in \href{https://github.com/dsb-lab/SignalingOscillators}{https://github.com/dsb-lab/SignalingOscillators}, which includes the parameter values used here.}
\label{fig:fig2}
\end{figure*}

The relation between the Notch pathway and signaling oscillations is underpinned by the fact that one of the targets of Notch is Hes1, the core component of the segmentation clock \citep{lewis2003autoinhibition}.
Additionally, the Notch ligand Delta-like1 (Dll1) has been seen to oscillate in multiple tissues, including the presomitic mesoderm, pancreatic and neural progenitors cells \citep{Shimojo:2016vn,Seymour:2020tc} and, more recently, in myogenic cells \cite{zhang2021oscillations} (Fig. \ref{fig:fig2}A).
In the latter case, the authors show that Dll1 is repressed by Hes1 directly, which suggests that the Dll1 oscillations are a downstream effect of the well-known oscillatory character of Hes1.

Besides its direct repressing effect, Zhang et al also show \cite{zhang2021oscillations} that Hes1 inhibits Dll1 indirectly by repressing the myogenic factor MyoD, which activates Dll1 (Fig. \ref{fig:fig2}B).
Interestingly, the dynamics of MyoD determines whether a myogenic cell self-renew or differentiates: while sustained oscillations of MyoD in activated stem cells are necessary for self-renewal and maintenance of the stem cell pool, sustained MyoD expression leads to differentiation.
This is shown in the simulations of Fig. \ref{fig:fig2}C, in which Hes1 oscillations arise from the following circuit:
\begin{subequations}\label{eq:sys2}
\begin{align}
    \frac{dH}{dt} &= k_1 h - k_2  H  F - k_3 H\label{eq:sys2a} \\
    \frac{dh}{dt} &= \frac{k_4}{1+H^2} - k_5 h\label{eq:sys2b} \\
    \frac{dF}{dt} &= \frac{k_6}{1+H^2} - k_2 H F - k_7  F \label{eq:sys2c}
\end{align}
\end{subequations}
where $H$ and $h$ represent Hes1 protein and mRNA, respectively, and $F$ is an interacting factor introduced in early models of the circuit \cite{Hirata:2002uc, lahmann2019oscillations}.
The model shows that the delay introduced by $F$, together with the negative feedback acting through $h$, are enough to produce sustained oscillations of Hes1.
The oscillation is then propagated to the rest of the system by the repression of $H$ on MyoD and Dll1 (Fig. \ref{fig:fig2}B). 
Furthermore, since Dll1 activates Notch signaling (which in turn activates Hes1) in neighboring cells, an additional inter-cell feedback arises that couples the oscillations in neighboring cells, the time scale of which needs to be tuned to avoid oscillation quenching, as Zhang et al verify experimentally using Dll1 mutants \cite{zhang2021oscillations}. This brings us back to the somitogenesis clock, where Notch signaling was seen to synchronize the oscillations that arise in the presomitic mesoderm \citep{Riedel-Kruse:2007va}.

\section{Intermittent ERK oscillations under FGF4 stimulation}

\begin{figure}[htbp]
    \centering
    \hspace*{-0.5cm}
    \includegraphics[width=\linewidth]{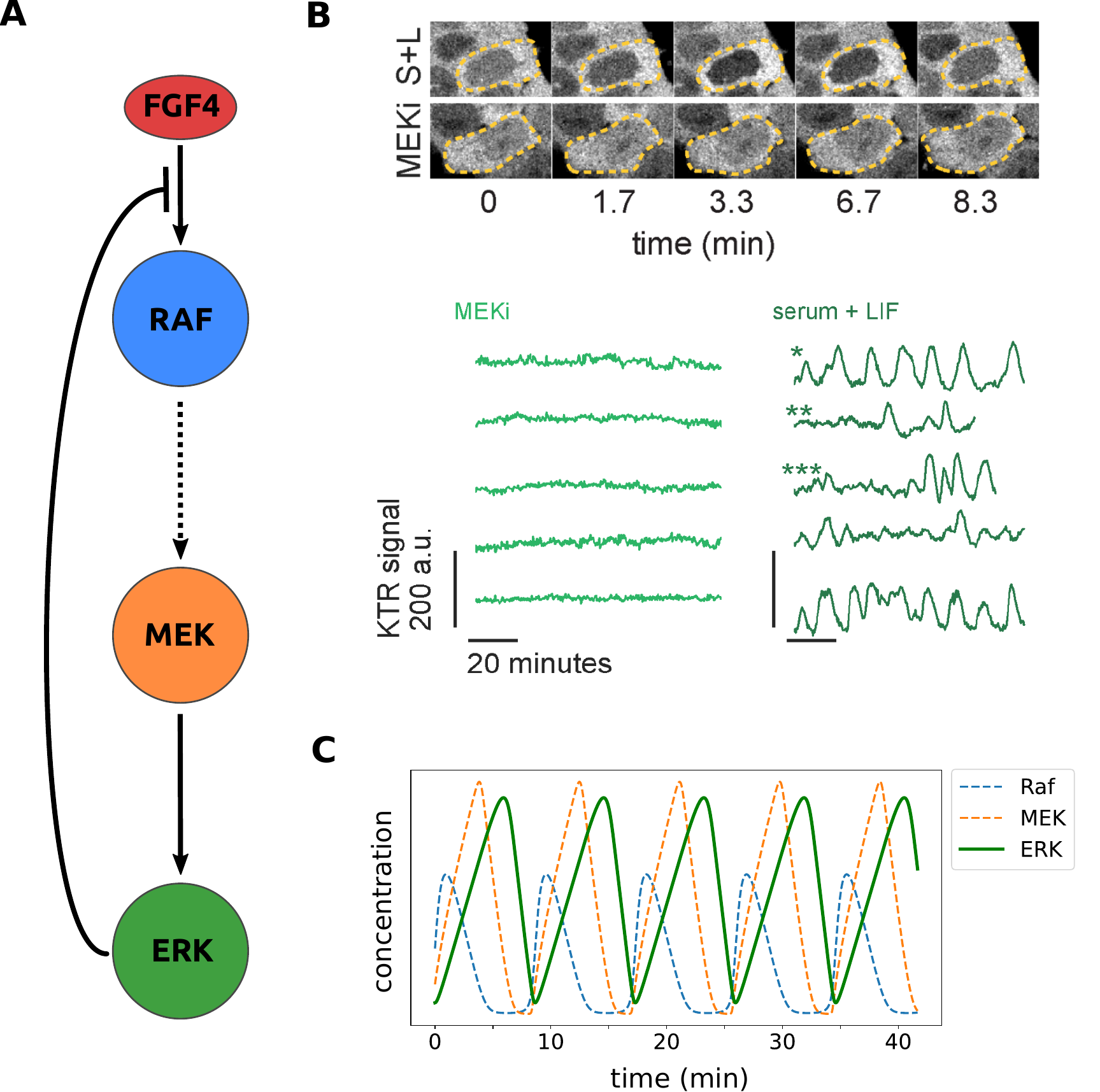}
    \caption{Oscillations in ERK activity mediated by negative feeback and ultrasensitivity. (\textbf{A}) Schematic representation of the circuit. The dashed line representes the ultrasensitive step.
    (\textbf{B}) Experimental observations of ERK oscillations in mouse ESCs, from \cite{raina2020intermittent}. Top: ERK-KTR expressing cells growing in serum + LIF without (top filmstrip) and with MEKi (bottom filmstrip). Dashed line indicates cell outlines. Bottom: time traces of the ERK activity of cells treated with a MEK inhibitor (left) or with serum + LIF (right).
    (\textbf{C}) Simulation of model (\ref{eq:sys3}). The values of the rest of parameters, together with the simulation code, can be found in
    \href{https://github.com/dsb-lab/SignalingOscillators}{https://github.com/dsb-lab/SignalingOscillators}.}	
    \label{fig:fig3}
\end{figure}

Our second example involves the extracellular regulated kinase (ERK). 
ERK is a classical mitogen-activated kinase (MAPK) module that, like Notch, is involved in a large number of cellular processes including differentiation, growth, proliferation, cell survival and apoptosis \cite{wortzel2011erk, gagliardi2021collective, simon2020live}. 
The ERK signaling pathway is a paracrine activated cascade by which an extracellular growth factor, such as the fibroblast growth factor (FGF), activates the Ras-Raf-MEK-ERK cascade (Fig \ref{fig:fig3}A). 
The cascade consists of a chain of serial phosphorylations that result in the double phosphorylation of ERK and its shuttling to the nucleus, where it affects the expression of more than 600 direct targets \cite{unal2017compendium}.
Even though the pathway is relatively direct, the dynamics of ERK localization and activity can be very complex: ERK is able to display a large variety of dynamic responses in different cellular contexts, even when presented with similar extracellular signals.
In particular, under sustained growth-factor stimulation, ERK can display sustained, pulsatile or oscillatory dynamics \cite{shankaran2009rapid,wilson2017tracing,li2019communication}.

Recently, it was shown that treating mouse embryonic stem cells (mESCs) with sustained concentrations of FGF4 results in oscillations of their ERK activity, which disappear in the presence of a MEK inhibitor \cite{raina2020intermittent} (Fig \ref{fig:fig3}B). 
The oscillations became more regular as the concentration of FGF4 is increased, although the shape of the pulses does not depend noticeably on the FGF4 levels.
Notably, ERK activity jumps between an oscillatory and non-oscillatory state over a physiological range of ligand concentrations. 
The role of these oscillations is still to be determined. 
It could merely be a mechanism of adaptation to keep the ERK response within physiological ranges, as proposed by the authors \cite{raina2020intermittent}.

Similar oscillations in the ERK signaling pathway have been identified in other studies. 
Stimulating mammalian epithelial cells with continuous epidermal growth factor (EGF), for instance, leads to oscillations of ERK nuclear translocation with a periodicity of 15 mins \cite{shankaran2009rapid}.
Increasing EGF levels reduce the amplitude of the oscillations while the pulse duration remains constant, similarly to the observations in mESCs described above. 
In contrast, other studies have shown that pulsing can be more stochastic than regular, and that growth-factor concentration can affect the frequency of the oscillation \cite{albeck2013frequency,aoki2013stochastic}.

The molecular mechanism of the oscillations observed by Raina et al \cite{raina2020intermittent} is still unclear.
A natural candidate is negative feedback, which is commonly found affecting ERK signaling \cite{lake2016negative}. 
Given that the time scale of the oscillation is of the order of 5 to 10 minutes, it is possible that the feedback is acting at the level of ERK phosphorilation/dephosphorilation, as suggested previously \cite{shankaran2009rapid, wilson2017tracing,kholodenko2000negative}.
The system depicted in figure \ref{fig:fig3}A is a simplification of the cascade, where we only consider the activated kinases and the feedback affects directly the input of the cascade.
Its dynamics can be represented by a model such as:
\begin{subequations}\label{eq:sys3}
\begin{align}
	\frac{dR}{dt} &= \beta_{R1} \frac{K_{R1}^n}{K_{R1}^n+E^n} \frac{F}{K_{R2}+F} - \frac{\beta_{R2}R}{K_{R3} + R} \\
	\frac{dM}{dt} &= \beta_{M1}\frac{R^m}{K_{M1}^m + R^m} - \frac{\beta_{M2}M}{K_{M2} + M} \\
	\frac{dE}{dt} &= \beta_{E1} \frac{M^p}{K_{E1}^p + M^p} - \frac{\beta_{E2}E}{E_{E2} + E}
\end{align}
\end{subequations}
Simulations of this system (\ref{eq:sys3}) show that in the presence of FGF (represented in the model by $F$) and large enough ultrasensitivity ($m\gg1$) in the activation of MEK ($M$) by RAF ($R$), ERK ($E$) displays sustained oscillations (Fig. \ref{fig:fig3}C).
These results are not sufficient, however, to explain all the characteristics that ERK activity displays in the presence of FGF4, in particular their intermittent character \cite{raina2020intermittent}. 
Potential extensions of the model (both conceptually and mathematically) could include the existence of multiple feedback loops acting at different steps of the cascade and at different timescales \cite{lake2016negative}. 

\section{ERK mechanochemical wave}
\begin{figure*}[htbp]
    \centering
    \includegraphics[width=\linewidth]{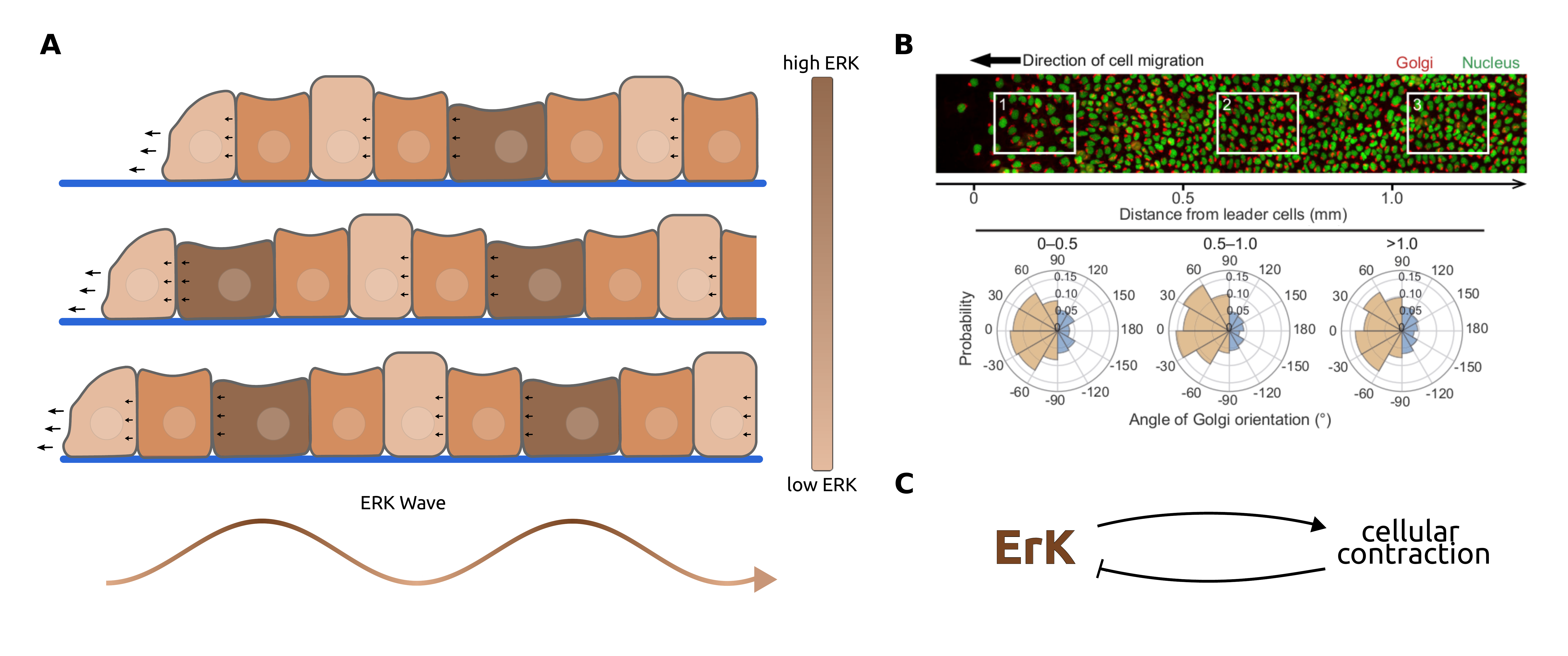}
    \caption{Mechanochemical feedbacks mediated by ERK drive collective motion. (\textbf{A}) Schematic representation of the mechanochemical system. Black arrows represent direction of cell movement while brown arrow represents direction of ERK wave. Time runs from top to bottom. (\textbf{B}) Immunofluorescence images of the Golgi apparatus (GM130) and the nucleus (EKAREV-NLS) in MDCK cells at 21 h after migration. Adapted from Hino et al \cite{hino2020erk}.(\textbf{C}) Schematic representation of the feedback circuit}
    \label{fig:fig4}
\end{figure*}

The negative feedbacks that we have considered so far have been purely biochemical.
In this last example, we are going to focus on signaling oscillations mediated by a delayed negative feedback that involves an elegant combination of biochemical and mechanical signals. 
Mechanical inputs are important factors in many different biological processes, such as the collective migration of the cells forming monolayers of epithelial tissue (Fig. \ref{fig:fig4}A,B) \cite{trepat2009physical, aoki2017propagating}. 
Waves of cell motion appear in contexts such as wound healing. 
Both in vivo and in vitro studies have shown that waves of ERK activation propagate counter to the direction of cell migration and cell polarization \cite{aoki2017propagating}. 
ERK waves are responsible for the directionality of the migration, by orienting the cells in the direction opposite to their propagation. 
In the absence of an external cue (e.g. a wound), there is no collective net movement of the tissue, even though there might be stochastic cellular motion. 
In the context of wound healing, on the other hand, the presence of stress, caused by the loss of congruency in the tissue, serves as a continuous input of mechanical force into the monolayer. 
Such continuous pulling produces a wave of ERK that propagates in the opposite direction of net cell movement \cite{hino2020erk, boocock2021theory}. 

What causes these ERK waves?
During wound healing each cell following the leader cell undergoes a series of activations and inactivations of ERK activity (Fig.~\ref{fig:fig4}A).
The mechanism by which cells transduce the mechanical intercellular interaction into an intracellular biochemical response was unknown until very recently.
Hino and colleagues \cite{hino2020erk} have demonstrated that the mechanical pulling of cells by their neighbors activates ERK. 
This increase in ERK activity then triggers the contraction of the cell, which in turns removes the original mechanical activation of ERK.
The resulting oscillation in cell activity is then propagated mechanically along the tissue.
The presence of such mechanochemical feedback (Fig. \ref{fig:fig4}C) is an ideal example of how the simple scheme of a delayed negative feedback can drive complex functions like collective polarization of cells to migrate in a specific direction.
Boocock et al have \cite{boocock2021theory} have proposed a viable model of collective motion based on the above-mentioned scheme (Fig. \ref{fig:fig4}C) of delayed feedback between ERK activity and cellular contraction. 
Such simple but elegant mechanism is enough to reproduce the intricacies of both stochastic and directed collective movements in monolayers of epithelial tissue.

\section{Discussion}
Cells are constantly receiving signals coming from their environment, and they need to decide how to react appropriately to the information (commonly dynamical, frequently unpredictable) contained in these signals. 
The signaling pathways involved in cellular communication and decision are scarce compared to the amount of possible environmental situations that a cell can face.
To solve this underactuation problem, cells have evolved sophisticated communication codes \cite{li2019communication} through which information is actively processed.

One way in which external signals can be ``multiplexed'' within a number of signaling systems is by using dynamics \citep{Gabalda-Sagarra:2018uv}.
In this approach, different dynamical regimes of a regulatory factor (in the limiting case oscillations versus stationary behavior) can code for different situations, and thereby for different reactions that the cell must follow.
Here we have seen an example of this phenomenon in the case of tissue maintenance in muscle stem cells (where oscillations of Notch pathway elements determine the fate of those cells).
We have also seen that not only adult stem cells, but embryonic stem cells as well, exhibit signaling oscillations (in this case of the ERK pathway).
The biological function of these oscillations is however not yet clear.

Both of the above-mentioned instances of signaling oscillations seem to be based on delayed negative feedback.
While the feedback in those cases is purely biochemical, the importance of mechanical interactions in tissues makes it reasonable to expect that mechanics might also be involved in feedback loops that could underlie oscillatory behavior.
This is the case of the third example considered here, involving collective cell motion in epithelial tissue.
This situation also involves ERK activation, but here mechanical regulation is an essential part of the feedback.
Furthermore, mechanics naturally leads to the spatial propagation of the oscillations in the form of waves of ERK activity, which regulate the cell motion.

These are just a few recent examples that show the relevance of oscillations in cell signaling systems.
More work is needed to identify further regulatory processes driven by oscillatory dynamics, which will remain hidden unless single-cell time-resolved measurements are systematically performed in cell-biology studies.

\section{Acknowledgments}
We acknowledge financial support from the Spanish Ministry of Science and Innovation and FEDER (grant PGC2018-101251-B-I00), by the ``Maria de Maeztu'' Programme for Units of Excellence in R\&D (grant CEX2018-000792-M), and by the Generalitat de Catalunya (ICREA Academia programme). P.C.-G. is supported by a FPI doctoral fellowship from the Spanish Ministry of Science and Innovation (reference XXXXXX).


\bibliography{signosc_preprint.bib}

\end{document}